\documentclass[aps,twocolumn,showpacs,pra,twoside,amssymb,amsmath]{revtex4}
\usepackage{amssymb}
\usepackage{graphicx}
\usepackage{amsmath}
\usepackage{colordvi}
\usepackage{bbm}

\newcommand{\eins}{\ensuremath{\mathbbm 1}}

\begin{document}

\title{Entanglement detection beyond the computable cross-norm or realignment criterion}
\author{Cheng-Jie Zhang}
\email{zhangcj@mail.ustc.edu.cn}
\author{Yong-Sheng Zhang}
\email{yshzhang@ustc.edu.cn}
\author{Shun Zhang}
\author{Guang-Can Guo}
\affiliation{Key Laboratory of Quantum Information, University of
Science and Technology of China, Hefei, Anhui 230026, People's
Republic of China}

\begin{abstract}
Separability problem, to decide whether a given state is entangled
or not, is a fundamental problem in quantum information theory. We
propose a powerful and computationally simple separability
criterion, which allows us to detect the entanglement of many bound
entangled states. The criterion is strictly stronger than the
criterion based on Bloch representations, the computable cross-norm
or realignment criterion and its optimal nonlinear entanglement
witnesses. Furthermore, this criterion can be generalized to an
analogue of permutation separability criteria in the even-partite
systems.
\end{abstract}


\maketitle

Entanglement is one of the most fascinating features of quantum
theory that has puzzled generations of physicist. While initially
the discussion was mainly driven by conceptual and philosophical
consideration \cite{Bell}, in recent years the focus has shifted to
mathematical aspects and practical applications. It was realized
that entanglement is an essential resource in quantum information
and acts an important role in many other physical phenomenon e.g.
quantum phase transition \cite{QPT}. Therefore, the detection and
quantification of entanglement became fundamental problems in
quantum information theory \cite{werner}. However, entanglement is
not yet fully understood and it is a challenging task and remains an
open question to decide whether a given state is entangled or not
despite a great deal of effort in the past years
\cite{review1,Peres,CCN1,CCN,permutation1,permutation2,dV,
LUR1,witness1,nonlinear,optimize,Yu,CM}.

Completely solving the separability problem is far away from us,
this is in fact a NP-hard problem as proved in \cite{Gurvits}.
Nevertheless, there are still many efficient conditions for
entanglement such as the partial transposition criterion
\cite{Peres}, the computable cross-norm or realignment (CCNR)
criterion \cite{CCN1,CCN}, the permutation separability criteria
\cite{permutation1,permutation2}, the criterion based on Bloch
representations \cite{dV}, local uncertainty relations \cite{LUR1},
entanglement witnesses \cite{witness1} and the covariance matrices
(CM) approach \cite{CM}, etc..
The CCNR criterion is only a necessary condition for arbitrary
dimensional systems. However, it can detect many bound entangled
states where the partial transposition criterion fails.
Recently, G\"uhne \textit{et al.} proposed its nonlinear
entanglement witness based on local uncertainty relations
\cite{nonlinear}. The nonlinear entanglement witness is strictly
stronger than the original criterion.

In this paper, we propose a practical criterion, based on
$\rho-\rho_{A}\otimes\rho_{B}$ which has similar properties as
covariance matrices \cite{CM}. The criterion is strictly stronger
than the dV criterion (i.e. the criterion based on Bloch
representations), the CCNR criterion and its optimal nonlinear
entanglement witnesses. Then we apply our criterion of separability
to a bound entangled state with white noise. Finally, we generalize
this criterion to multipartite entanglement and propose an analogue
of permutation separability criteria in even-partite systems. It is
worth noticing that our method proposed in this paper may be used to
improve many other separability criteria.

\textit{Bipartite systems.--} Before embarking on our criteria, it
is worth introducing the CCNR criterion and its nonlinear witnesses.
The CCNR criterion states that if $\rho$ is separable, the following
inequality must be hold \cite{CCN},
\begin{equation}\label{CCN}
    \|\mathcal{R}(\rho)\|\leq1,
\end{equation}
where $\|\cdot\|$ stands for the trace norm (i.e. the sum of the
singular values). The realignment operation $\mathcal{R}(A\otimes
B)=|A\rangle\langle B^{*}|$, with scalar product $\langle
B|A\rangle=\mathrm{tr}(B^{\dag}A)$ in Hilbert Schmidt space of
operators, and for a general operator it is given by linearity
expanding in a product basis \cite{CCN1,CCN,permutation1,Appendix}.
Refs. \cite{nonlinear,optimize} put forward its nonlinear witnesses
and their optimal form, respectively,
\begin{eqnarray}
    \mathcal{F}(\rho)=1-\sum_{k}\langle G_{k}^{A}\otimes
    G_{k}^{B}\rangle-\frac{1}{2}\sum_{k}\langle G_{k}^{A}\otimes\eins-\eins\otimes
    G_{k}^{B}\rangle^{2},\label{nonlinear}\\
\mathcal{F}_{opt}(\rho)=1-\|\tau\|-(\mathrm{Tr}\rho_{A}^{2}+
\mathrm{Tr}\rho_{B}^{2})/2,\label{optimal}
\end{eqnarray}
where $\{G_{k}^{A}\}$, $\{G_{k}^{B}\}$ are complete sets of local
orthogonal bases \cite{Yu} for subsystems $A$ and $B$ respectively,
and $\tau$ is defined as $\tau_{lm}=\langle G_{l}^{A}\otimes
G_{m}^{B}\rangle-\langle
G_{l}^{A}\otimes\eins\rangle\langle\eins\otimes G_{m}^{B}\rangle$.
For separable states $\mathcal{F}(\rho)\geq0$ and
$\mathcal{F}_{opt}(\rho)\geq0$ hold. Conversely, if any state
violates one of the three inequalities, it is indeed entangled.
Actually, $\|\tau\|$ can be expressed as
$\|\mathcal{R}(\rho-\rho_{A}\otimes\rho_{B})\|$, which will be
proved in Proposition 1. In the following, we will propose a
separability criterion. It is a slightly modified and therefore
improved version of the optimal nonlinear witness Eq.
(\ref{optimal}).

\textit{Theorem 1.} If a bipartite density matrix $\rho$ is
separable, then the following inequality holds,
\begin{equation}\label{th1}
    \|\mathcal{R}(\rho-\rho_{A}\otimes\rho_{B})\|\leq
    \sqrt{(1-\mathrm{Tr}\rho_{A}^{2})(1-\mathrm{Tr}\rho_{B}^{2})},
\end{equation}
where $\rho_{A}$ and $\rho_{B}$ are reduced density matrices for
subsystems $A$ and $B$.

\textit{Proof.--} A separable bipartite density matrix $\rho$ can be
written as $\rho=\sum_{i}p_{i}\rho_{i}^{A}\otimes\rho_{i}^{B}$, and
its reduced density matrices are
$\rho_{A}=\sum_{i}p_{i}\rho_{i}^{A}$,
$\rho_{B}=\sum_{i}p_{i}\rho_{i}^{B}$, where $\{p_{i}\}$ is a
probability distribution and the $\rho_{i}^{A}$, $\rho_{i}^{B}$ are
pure states describing subsystems $A$ and $B$, respectively.

It is easy to conclude that
$\rho-\rho_{A}\otimes\rho_{B}=\frac{1}{2}\sum_{i,j}p_{i}p_{j}
(\rho_{i}^{A}-\rho_{j}^{A})\otimes(\rho_{i}^{B}-\rho_{j}^{B})$. We
have reviewed the realignment operation $\mathcal{R}$ in
\cite{Appendix}. Therefore,
\begin{eqnarray}
&&\|\mathcal{R}(\rho-\rho_{A}\otimes\rho_{B})\|\nonumber\\
&=&\frac{1}{2}\|\sum_{i,j}p_{i}p_{j}(|\rho_{i}^{A}\rangle-|\rho_{j}^{A}\rangle)(\langle(\rho_{i}^{B})^{*}|-\langle(\rho_{j}^{B})^{*}|)\|\nonumber\\
&\leq&\frac{1}{2}\sum_{i,j}p_{i}p_{j}\|(|\rho_{i}^{A}\rangle-|\rho_{j}^{A}\rangle)(\langle(\rho_{i}^{B})^{*}|-\langle(\rho_{j}^{B})^{*}|)\|\nonumber\\
&=&\sum_{i,j}(\sqrt{p_{i}p_{j}}\sqrt{1-\mathrm{Tr}\rho_{i}^{A}\rho_{j}^{A}})(\sqrt{p_{i}p_{j}}\sqrt{1-\mathrm{Tr}(\rho_{i}^{B}\rho_{j}^{B})^{*}})\nonumber\\
&\leq&\sqrt{[\sum_{i,j}p_{i}p_{j}(1-\mathrm{Tr}\rho_{i}^{A}\rho_{j}^{A})][\sum_{i,j}p_{i}p_{j}(1-\mathrm{Tr}\rho_{i}^{B}\rho_{j}^{B})]}\nonumber\\
&=&\sqrt{(1-\mathrm{Tr}\rho_{A}^{2})(1-\mathrm{Tr}\rho_{B}^{2})},
\end{eqnarray}
where we have used
$\mathrm{Tr}(\rho_{i}^{B}\rho_{j}^{B})^{*}=\mathrm{Tr}(\rho_{i}^{B}\rho_{j}^{B})^{T}=\mathrm{Tr}\rho_{i}^{B}\rho_{j}^{B}$.
The first inequality holds due to the convex property of the trace
norm and the second one holds by applying the Cauchy-Schwarz
inequality. \hfill $\square$

Obviously, Theorem 1 has a similar form of the CCNR criterion, using
$\rho-\rho_{A}\otimes\rho_{B}$ and
$\sqrt{(1-\mathrm{Tr}\rho_{A}^{2})(1-\mathrm{Tr}\rho_{B}^{2})}$
instead of $\rho$ and 1, respectively. It suggests that using
$\rho-\rho_{A}\otimes\rho_{B}$ we can obtain some new separability
criteria. In the following, it will be shown that Theorem 1 is
strictly stronger than the CCNR criterion and its nonlinear
witnesses (\ref{nonlinear}), with an example and a strict proof.

\begin{figure}
\begin{center}
\includegraphics[scale=0.4]{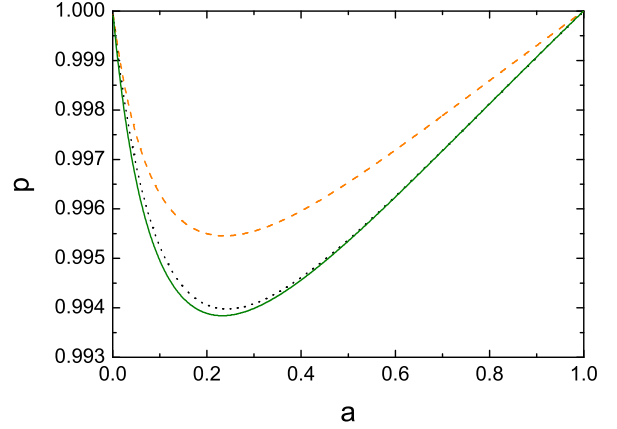}
\caption{(color online). Detecting the entanglement of Horodecki
$3\times3$ bound entangled state with white noise. The regions above
the curves can be detected as entangled states by the CCNR criterion
(dashed line), its optimal nonlinear witness (dotted line), and
Theorem 1 (solid line), respectively.}\label{1}
\end{center}
\end{figure}

\textit{Example 1.--} Pawe{\l} Horodecki introduced a $3\times3$
bound entangled state in Ref. \cite{bound}, and the density matrix
$\rho$ is real and symmetric,
\begin{equation}
\rho ={\frac{1}{8a+1}}\left(
\begin{array}{ccccccccc}
a & 0 & 0 & 0 & a & 0 & 0 & 0 & a \\
0 & a & 0 & 0 & 0 & 0 & 0 & 0 & 0 \\
0 & 0 & a & 0 & 0 & 0 & 0 & 0 & 0 \\
0 & 0 & 0 & a & 0 & 0 & 0 & 0 & 0 \\
a & 0 & 0 & 0 & a & 0 & 0 & 0 & a \\
0 & 0 & 0 & 0 & 0 & a & 0 & 0 & 0 \\
0 & 0 & 0 & 0 & 0 & 0 & {\frac{1+a}{2}} & 0 & {\frac{\sqrt{1-a^{2}}}{2}} \\
0 & 0 & 0 & 0 & 0 & 0 & 0 & a & 0 \\
a & 0 & 0 & 0 & a & 0 & {\frac{\sqrt{1-a^{2}}}{2}} & 0 &
{\frac{1+a}{2}}
\end{array}
\right), \label{rho}
\end{equation}
where $0<a<1$. Let us consider a mixture of this state with white
noise,
\begin{equation}\label{mix}
    \rho(p)=p\rho+(1-p)\frac{\eins}{9},
\end{equation}
and show the curves $1-\|\mathcal{R}(\rho)\|=0$,
$1-\|\tau\|-(\mathrm{Tr}\rho_{A}^{2}+ \mathrm{Tr}\rho_{B}^{2})/2=0$,
$\sqrt{(1-\mathrm{Tr}\rho_{A}^{2})(1-\mathrm{Tr}\rho_{B}^{2})}
    -\|\mathcal{R}(\rho-\rho_{A}\otimes\rho_{B})\|=0$ with respect to
the CCNR criterion, its optimal nonlinear witness, and Theorem 1 in
Fig. \ref{1}.

In Ref. \cite{CCN}, it is found that the state $\rho(p)$ still has
entanglement when $p=0.9955$, $a=0.236$, using the CCNR criterion.
According to Theorem 1, one can obtain an upper bound $p=0.9939$,
$a=0.232$ for $\rho(p)$ which is still entangled. Furthermore,
states $\rho(p)$ which can be detected by the CCNR criterion or its
nonlinear witnesses also violate Theorem 1 (see Fig. \ref{1}).
Proposition 1 will provide a strict proof.

\textit{Proposition 1.} Any state which can be detected by the CCNR
criterion or its nonlinear witnesses (\ref{nonlinear}) also violates
Theorem 1.

\textit{Proof.--} It is worth noticing that Eq. (\ref{optimal}) is
equivalent to
$\mathcal{F}_{opt}(\rho)=1-\|\mathcal{R}(\rho-\rho_{A}\otimes\rho_{B})\|-(\mathrm{Tr}\rho_{A}^{2}+
\mathrm{Tr}\rho_{B}^{2})/2$, i.e., the sum of singular values of
matrix $\tau$ is equal to the trace norm of
$\mathcal{R}(\rho-\rho_{A}\otimes\rho_{B})$. Note that
$\sum_{l,m}\tau_{lm}G_{l}\otimes G_{m} =\sum_{l,m}\langle
G_{l}\otimes G_{m}\rangle G_{l}\otimes G_{m} -(\sum_{l}\langle
G_{l}\otimes\eins\rangle
G_{l}\otimes\eins)(\sum_{m}\langle\eins\otimes
G_{m}\rangle\eins\otimes G_{m}) =\rho-\rho_{A}\otimes\rho_{B}$.
Therefore, $\|\mathcal{R}(\rho-\rho_{A}\otimes\rho_{B})\|
=\|\sum_{l,m}\tau_{lm}|G_{l}\rangle\langle G_{m}^{*}|\| =\|\tau\|$,
where $\langle
G_{l^{'}}|G_{l}\rangle=\mathrm{Tr}(G_{l^{'}}^{\dag}G_{l})=\delta_{ll^{'}}$.
Moreover, Theorem 1 can be written as a nonlinear witness,
$\mathcal{G}=\sqrt{(1-\mathrm{Tr}\rho_{A}^{2})(1-\mathrm{Tr}\rho_{B}^{2})}
-\|\mathcal{R}(\rho-\rho_{A}\otimes\rho_{B})\|$.

Due to
$\sqrt{(1-\mathrm{Tr}\rho_{A}^{2})(1-\mathrm{Tr}\rho_{B}^{2})}\leq1-(\mathrm{Tr}\rho_{A}^{2}+
\mathrm{Tr}\rho_{B}^{2})/2$, it can be concluded that Theorem 1 is
strictly stronger than the optimized nonlinear witness Eq.
(\ref{optimal}). Since Eq. (\ref{optimal}) is not only a nonlinear
form of Eq. (\ref{CCN}) but also an optimal form of Eq.
(\ref{nonlinear}), it is strictly stronger than Eqs. (\ref{CCN}) and
(\ref{nonlinear}) \cite{nonlinear,optimize}. Thus, Proposition 1
holds. \hfill $\square$

\textit{Example 2.--} Let us consider a noisy singlet state
introduced in Ref. \cite{nonlinear}, $\rho=p|\psi_{s}\rangle\langle
\psi_{s}|+(1-p)\rho_{sep}$, where
$|\psi_{s}\rangle=(|01\rangle-|10\rangle)/\sqrt{2}$ and
$\rho_{sep}=2/3|00\rangle\langle00|+1/3|01\rangle\langle01|$.
Actually, the state is entangled for any $p>0$ \cite{nonlinear}. The
CCNR criterion and its optimal nonlinear witness can detect the
entanglement for all $p>0.292$ and $p>0.25$, respectively. Using
Theorem 1, one finds that the state still has entanglement when
$p>0.221$. One might expect Theorem 1 to be a necessary and
sufficient condition for entanglement in two-qubit system.
Unfortunately, this is not the case. However, considering the
enhancement with local filtering operations, Theorem 1 becomes a
necessary and sufficient condition for two qubits \cite{CM}.


\textit{Proposition 2.} Theorem 1 is strictly stronger than the dV
criterion.

\textit{Proof.--} For $M\times N$ bipartite systems, the Bloch
representation can be written as
$\rho=(\eins_{M}\otimes\eins_{N}+\sum_{i}r_{i}\lambda_{i}^{A}\otimes\eins_{N}
+\sum_{j}s_{j}\eins_{M}\otimes\lambda_{j}^{B}+\sum_{ij}t_{ij}\lambda_{i}^{A}\otimes\lambda_{j}^{B})/MN$,
where $\{\lambda_{i}^{A}\}$ and $\{\lambda_{j}^{B}\}$ denote the
generators of SU(M) and SU(N). The coefficients $t_{ij}$, $r_{i}$,
$s_{j}$ form the real matrix $\mathrm{T}$, and column vectors
$\mathrm{r}$, $\mathrm{s}$, respectively. The dV criterion states
that if a bipartite state is separable then
$\|\mathrm{T}\|\leq\sqrt{MN(M-1)(N-1)/4}$ must hold \cite{dV}.
Notice that Theorem 1 is equivalent to
$\|\mathrm{T}-\mathrm{r}\cdot\mathrm{s}^{T}\|\leq\sqrt{\frac{MN}{4}(M-1-\frac{2\sum_{i}r_{i}^{2}}{M})(N-1-\frac{2\sum_{j}s_{j}^{2}}{N})}$.
With the help of the triangle inequality of trace norm, the
left-hand side (LHS) can be bounded as
$\mathrm{LHS}=\|\mathrm{T}-\mathrm{r}\cdot\mathrm{s}^{T}\|\geq\|\mathrm{T}\|-\|\mathrm{r}\cdot\mathrm{s}^{T}\|=\|\mathrm{T}\|-\sqrt{(\sum_{i}r_{i}^{2})(\sum_{j}s_{j}^{2})}$.
For the right-hand side (RHS), we find
$\mathrm{RHS}=\sqrt{\frac{MN}{4}(M-1-\frac{2\sum_{i}r_{i}^{2}}{M})(N-1-\frac{2\sum_{j}s_{j}^{2}}{N})}
\leq\sqrt{\frac{MN}{4}(M-1)(N-1)}-\sqrt{(\sum_{i}r_{i}^{2})(\sum_{j}s_{j}^{2})}$.
From $\mathrm{LHS}\leq \mathrm{RHS}$, we can conclude that
$\|\mathrm{T}\|\leq\sqrt{MN(M-1)(N-1)/4}$ holds. Thus, any state
which satisfies Theorem 1 must satisfy the dV criterion as well,
i.e., Theorem 1 is strictly stronger than the dV criterion. \hfill
$\square$

It was pointed out in Ref. \cite{Referee}, if one considers the
enhancement of separability criteria with local filtering operations
introduced in Ref. \cite{CM}, Theorem 1 reduces to the dV criterion
\cite{Referee}. By constructive algorithms, states can be
transformed (preserving entanglement or separability) to a filter
normal form (FNF) or arbitrarily close to it \cite{CM}. For states
in the FNF, $\langle G_i\otimes\eins\rangle=\langle\eins\otimes
G_i\rangle=0$ holds. Therefore, for these states Theorem 1 is
equivalent to the dV criterion. Hence, Theorem 1 is not expected to
improve our entanglement detection capability if one is able to
enhance other criteria with local filters. However, there are still
some advantages \cite{Referee}. Firstly, separability criteria under
filtering require numerical algorithms, which might pose problems as
the dimensionality increases, while Theorem 1 is completely
analytical. Secondly, the results rely on interesting, original and
relatively simple ideas which might be used to improve other
criteria. Finally, we will generalize Theorem 1 to the multipartite
setting and derive a criterion (Theorem 2) for states with an even
number of subsystems. To compare Theorem 1 with inequality (8) in
\cite{CM} in the FNF, one can conclude that for $d_{A}=d_{B}$ they
coincide, where $d_{A}$ ($d_{B}$) is the dimension of subsystem A
(B). If $|d_{B}-d_{A}|$ is small, Theorem 1 is slightly better than
inequality (8) in \cite{CM}, if $|d_{B}-d_{A}|$ is large, inequality
(8) in \cite{CM} is drastically better than Theorem 1 \cite{CM}.

The transformation $\rho\rightarrow\rho-\rho_{A}\otimes\rho_{B}$
used in Theorem 1 can also be used to obtain a criterion which is
similar to the partial transposition criterion.

\textit{Proposition 3.} If a bipartite density matrix $\rho$ is
separable, then the following inequality holds,
\begin{equation}\label{pr2}
    \|(\rho-\rho_{A}\otimes \rho_{B})^{T_{B}}\|\leq
2\sqrt{(1-\mathrm{Tr}\rho_{A}^{2})(1-\mathrm{Tr}\rho_{B}^{2})},
\end{equation}
where $T_{B}$ stands for a partial transpose with respect to the
subsystem $B$.

\textit{Proof.--} For a separable bipartite density matrix $\rho$,
it can be concluded that $
\|(\rho-\rho_{A}\otimes\rho_{B})^{T_{B}}\|
\leq\frac{1}{2}\sum_{i,j}p_{i}p_{j}
\|\rho_{i}^{A}-\rho_{j}^{A}\|\cdot\|\rho_{i}^{B}-\rho_{j}^{B}\|,\nonumber
$ where we used the equalities $\|A\otimes B\|=\|A\|\cdot\|B\|$
\cite{horn2} and $\|B^{T}\|=\|B\|$. Notice that
$\mathrm{rank}(\rho_{i}^{A}-\rho_{j}^{A})\leq2$,
$\mathrm{Tr}(\rho_{i}^{A}-\rho_{j}^{A})=0$ and
$(\rho_{i}^{A}-\rho_{j}^{A})^{\dag} =\rho_{i}^{A}-\rho_{j}^{A}$.
Thus, the eigenvalues of $\rho_{i}^{A}-\rho_{j}^{A}$ can be written
as $\lambda$, $-\lambda$ ($\lambda\geq0$) and the singular values
are $\lambda$, $\lambda$. Due to
$\mathrm{Tr}(\rho_{i}^{A}-\rho_{j}^{A})^{2}=2\lambda^{2}$, it is
obtained that
    $\|\rho_{i}^{A}-\rho_{j}^{A}\|=\sqrt{2\mathrm{Tr}(\rho_{i}^{A}-\rho_{j}^{A})^{2}}
    =2\sqrt{1-\mathrm{Tr}\rho_{i}^{A}\rho_{j}^{A}}.$
Similarly,
$\|\rho_{i}^{B}-\rho_{j}^{B}\|=2\sqrt{1-\mathrm{Tr}\rho_{i}^{B}\rho_{j}^{B}}$
can be gotten. Thus, we have
$\|(\rho-\rho_{A}\otimes\rho_{B})^{T_{B}}\|\leq2\sqrt{(1-\mathrm{Tr}\rho_{A}^{2})(1-\mathrm{Tr}\rho_{B}^{2})}$
with the Cauchy-Schwarz inequality. \hfill $\square$

Actually, there is a tiny difference between Theorem 1 and
Proposition 3. Right hand side of Eq. (\ref{pr2}) is exactly two
times as large as the one of Eq. (\ref{th1}). However, the
coefficient 2 cannot be replaced with a smaller number. For example,
when the separable state
$\rho=(|00\rangle\langle00|+|11\rangle\langle11|)/2$ is substituted
into inequality (\ref{pr2}), the equal sign holds. It is one of the
reasons that Proposition 3 is not as strong as Theorem 1. Consider
Example 2, it can only detect entanglement for $p>0.65$.

\textit{Multipartite systems.--} Theorem 1 and Proposition 3 can be
generalized to even-partite systems. A mixed state of an $N$-partite
system is defined to be separable if it can be represented in the
form $\rho=\sum_{i}p_{i}\rho_{i}
^{1}\otimes\rho_{i}^{2}\otimes\cdots\otimes\rho_{i}^{N}$, where
$\{p_{i}\}$ is a probability distribution and
$\rho_{i}^{1},\cdots,\rho_{i}^{N}$ are pure states of subsystems.
When $N$ is an even number, there are two different classes of
bipartite partitions $\mathcal{P_{I}}$ and $\mathcal{P_{II}}$
introduced in \cite{Cai}. $\mathcal{P_{I}}$ denotes that both sides
of bipartite partition contain odd number of parties, and
$\mathcal{P_{II}}$ means even-number parties in each side
\cite{note}. For instance,
$\mathcal{P_{I}}=\{\rho_{1}\otimes\rho_{234},\rho_{2}\otimes\rho_{134},
\rho_{3}\otimes\rho_{124},\rho_{4}\otimes\rho_{123}\}$ and
$\mathcal{P_{II}}=
\{\rho,\rho_{12}\otimes\rho_{34},\rho_{13}\otimes\rho_{24},\rho_{14}\otimes\rho_{23}\}$
\cite{note2} when $N=4$. An operator of their linear combination can
be defined
\begin{equation}\label{}
    \Delta\rho=\frac{1}{2^{N-2}}(\mathcal{Q_{II}}-\mathcal{Q_{I}}),
\end{equation}
where $\mathcal{Q_{II}}=\sum_{q\in\mathcal{P_{II}}}q$ and
$\mathcal{Q_{I}}=\sum_{p\in\mathcal{P_{I}}}p$. For $N=2$ and $4$,
$\Delta\rho=\rho-\rho_{1}\otimes\rho_{2}$ and
$\frac{1}{4}(\rho+\rho_{12}\otimes\rho_{34}+\rho_{13}\otimes\rho_{24}+\rho_{14}\otimes\rho_{23}-
\rho_{1}\otimes\rho_{234}-\rho_{2}\otimes\rho_{134}-\rho_{3}\otimes\rho_{124}-\rho_{4}\otimes\rho_{123})$,
respectively. In the following, we will present parallel criteria of
permutation separability criteria based on $\Delta\rho$. Recall the
permutation operation $\mathcal{R}_{(mn)}\otimes\eins$ introduced in
\cite{permutation1}, where $\mathcal{R}_{(mn)}$ acts on the $m$th
and $n$th parties while leaves untouched the rest subsystems.

\textit{Theorem 2 (General criteria).} If an $N$-partite density
matrix $\rho$ is separable ($N$ is an even number), then the
following inequalities
\begin{eqnarray}
    \|\mathcal{R}_{(mn)}\otimes\eins(\Delta\rho)\|\leq
    \min_{1\leq k\neq l\leq
N}\sqrt{(1-\mathrm{Tr}\rho_{k}^{2})(1-\mathrm{Tr}\rho_{l}^{2})},\label{R}\\
\|\mathcal{L}(\Delta\rho)\|\leq\min_{1\leq k\neq l\leq
N}\frac{1}{2^{\frac{N}{2}-1}}\sqrt{(1-\mathrm{Tr}\rho_{k}^{2})(1-\mathrm{Tr}
\rho_{l}^{2})}\label{L}
\end{eqnarray}
hold for separable states, where
$\mathcal{L}\equiv\mathcal{R}_{(i_{1}i_{2})}\otimes
\mathcal{R}_{(i_{3}i_{4})}\otimes\cdots\otimes\mathcal{R}_{(i_{N-1}i_{N})}$
and $\{i_{1},i_{2},\cdots,i_{N}\}$ is a rearrangement of
$\{1,2,\cdots,N\}$.

\textit{Proof.-} According to the proof of Theorem 1 and Proposition
3, it can be concluded that
$\|\mathcal{R}_{(mn)}\otimes\eins(\Delta\rho)\|
\leq\sum_{ij}p_{i}p_{j}\prod_{l=1}^{N}\sqrt{1-\mathrm{Tr}\rho_{i}^{l}\rho_{j}^{l}}
\leq\min_{1\leq k\neq l\leq
N}\sqrt{(1-\mathrm{Tr}\rho_{k}^{2})(1-\mathrm{Tr}\rho_{l}^{2})}, $
where we have used
$\Delta\rho=\frac{1}{2^{N-2}}(\mathcal{Q_{II}}-\mathcal{Q_{I}})
=\frac{1}{2^{N-1}}\sum_{ij}p_{i}p_{j}(\rho_{i}^{1}-\rho_{j}^{1})
\otimes(\rho_{i}^{2}-\rho_{j}^{2})\otimes\cdots\otimes(\rho_{i}^{N}-\rho_{j}^{N})$,
$\|(|\rho_{i}^{m}\rangle-|\rho_{j}^{m}\rangle)(\langle(\rho_{i}^{n})^{*}|
-\langle(\rho_{j}^{n})^{*}|)\|=2\sqrt{(1-\mathrm{Tr}\rho_{i}^{m}\rho_{j}^{m})
(1-\mathrm{Tr}\rho_{i}^{n}\rho_{j}^{n})}$ and
$\|\rho_{i}^{l}-\rho_{j}^{l}\|=2\sqrt{1-\mathrm{Tr}\rho_{i}^{l}\rho_{j}^{l}}$.
Inequality (\ref{L}) can also be proved with the same method.
\hfill $\square$


$\Delta\rho$ has a certain meaning. Notice that all of the criteria
presented in this paper can be viewed as parallel criteria of the
CCNR, partial transposition and permutation separability criteria
based on $\Delta\rho$, and they are independent on the original
criteria. It is considered that $\rho-\rho_{A}\otimes\rho_{B}$ and
the covariance matrix $\tau$ are of similar construction, where
$\mathcal{R}(\rho-\rho_{A}\otimes\rho_{B})$ having the same singular
values as $\tau$ can be viewed as evidence. Moreover, $\Delta\rho$
in multipartite systems seems to contain genuine entanglement
information in the sense of explanation
$S_{\mathcal{I}}-S_{\mathcal{II}}$ in Ref. \cite{Cai}. Therefore,
the operator $\Delta\rho$ can be viewed as removing some local and
separable information from $\rho$. We make a conjecture that
$\Delta\rho$ can also be used to obtain some other separable
criteria.

In conclusion, we have presented a more powerful separability
criterion, which is strictly stronger than the dV criterion, the
CCNR criterion and its optimal nonlinear entanglement witnesses .
The criterion is computationally simple and has been generalized in
even-partite systems. It is worth noting that many other
separability criteria may be improved with the method proposed in
this paper. It is an interesting open question whether our criteria
can be used to obtain lower bounds on the concurrence
\cite{concurrence,mintert,chen}, since Ref. \cite{chen} has derived
a lower bound of the concurrence based on the partial transposition
and CCNR criteria.

We thank Heng Fan and Otfried G\"uhne for helpful discussions, and
anonymous referees for pointing out that Theorem 1 reduces to the dV
criterion for states in the FNF and other suggestions. This work was
funded by the National Fundamental Research Program (Grant No.
2006CB921900), the National Natural Science Foundation of China
(Grants No. 10674127 and No. 60621064), the Innovation Funds from
the Chinese Academy of Sciences, Program for NCET.

\textit{Note added.--} After resubmission of the revised manuscript,
we became aware of a very recent preprint by Gittsovich \textit{et
al.} \cite{CM2}, which has proved that Theorem 1 is a corollary of
the CM criterion.

\textit{Appendix.--} Here, We review the realignment operation
$\mathcal{R}$ introduced by Chen and Wu \cite{CCN}. For each
$m\times n$ matrix $A=[a_{ij}]$, the vector $vec(A)$ is defined as
$vec(A)=[a_{11},\cdots,a_{m1},a_{12},\cdots,a_{m2},\cdots,a_{1n},
\cdots,a_{mn}]^{T}$. Suppose $Z$ is an $m\times m$ block matrix with
block size $n\times n$. The realignment $\mathcal{R}$ is defined as
$\mathcal{R}(Z)\equiv \left[
\begin{array}{llll}
vec(Z_{1,1})\cdots vec(Z_{m,1})\cdots vec(Z_{1,m})\cdots
vec(Z_{m,m})
\end{array}\right]^{T}$.
Therefore, a straightforward conclusion holds,
$\mathcal{R}(\rho_{A}\otimes \rho_{B})=vec(\rho_{A})\cdot
vec(\rho_{B})^{T} \equiv|\rho_{A}\rangle\langle\rho_{B}^{*}|$
\cite{CCN1,CCN,permutation1}, where
$\langle\rho_{B}|\rho_{A}\rangle=\mathrm{tr}(\rho_{B}^{\dag}\rho_{A})$
holds.


\begin{thebibliography}{99}
\bibitem{note} There are two differeces about the definitions between this paper
and Ref. \cite{Cai}. Firstly, $\mathcal{P_{I}}$ and
$\mathcal{P_{II}}$ in this paper are sets of direct products of
reduced density matrices under the bipartite partitions. Secondly,
$\rho$ is added into $\mathcal{P_{II}}$ (i.e., all of the parties
are in the same side of the bipartite partition in this case).
\bibitem{note2} The results of direct products should be represented in
the basis of $\rho$. For example, $\rho_{24}\otimes\rho_{13}$ is
equal to $\rho_{13}\otimes\rho_{24}$.
\end{thebibliography}
\end{document}